\documentclass[conference]{IEEEtran}

\usepackage{amsmath,amssymb,amsfonts}
\usepackage{graphicx}
\usepackage{textcomp}
\usepackage{xcolor}
\usepackage[numbers,sort&compress]{natbib}

\newcommand{\dataurl}{https://huggingface.co/datasets/DilushaChandrasiri/SriLanka-Bird-Diversity-Dataset}
\newcommand{\codeurl}{https://github.com/bird-diversity/bird_diversity_project.git}

\newcommand{\hf}[2]{\raisebox{-2.2pt}{\includegraphics[scale=0.09]{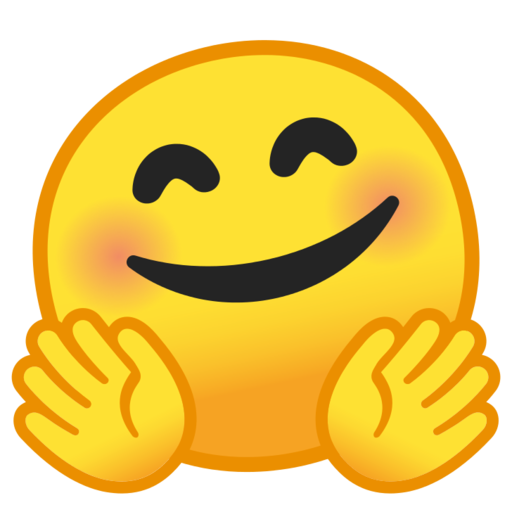}}~\href{#1}{\texttt{#2}}}
\newcommand{\gh}[2]{\raisebox{-2.2pt}{\includegraphics[scale=0.02]{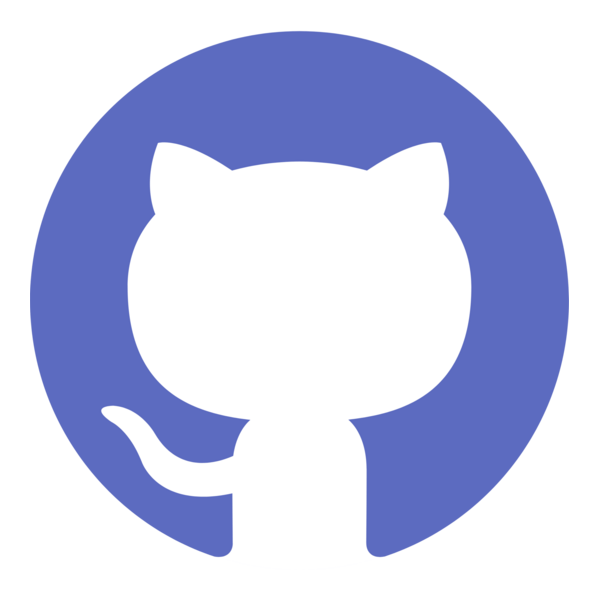}}~\href{#1}{\texttt{#2}}}

\usepackage{hyperref}
\hypersetup{
    colorlinks   = true,
    citecolor    = blue,
    urlcolor     = blue
}

\begin{document}
\title{How Environment and Urbanization Shape Bird Diversity in Sri Lanka}

\author{
\IEEEauthorblockN{%
Dilusha Chandrasiri,
Maneesha Herath,
Yasith Hewarathna,
Muditha Herath,
Gishan Bandara,\\
Madara Mendis,
Nathali Athukorala,
Nisansa de Silva,
Sandareka Wickramanayake}
\IEEEauthorblockA{Dept.\ of Computer Science \& Engineering, University of Moratuwa, Sri Lanka.\\
\texttt{\{dilushac.23, maneeshah.23, yasithh.23, mudithah.23, gishanb.23,}\\
\texttt{madara.21, nathalia, NisansaDds, sandarekaw\}@cse.mrt.ac.lk}
}
}

\maketitle

\begin{abstract}
This study presents a comprehensive analysis of bird diversity across Sri Lanka by integrating spatial, temporal, and environmental data. Bird observation records were combined with environmental variables, including weather conditions, air pollution, the Normalized Difference Vegetation Index (NDVI), land cover, elevation, and Artificial Light At Night (ALAN), and rigorously preprocessed to ensure data quality. Spatial analyses were conducted on multiple grid scales (2 km, 5 km, 10 km) to evaluate patterns in species richness while minimizing sampling bias through spatial thinning. Temporal trends were assessed using effort-corrected metrics including rarefied richness and occupancy rates to account for variations in observation effort over time. Environmental drivers of bird diversity were examined using multivariate statistical models, including Poisson Generalized Linear Models (GLMs) and correlation analyses, to identify key associations between ecological factors and species richness. Additionally, community structure, dominance patterns, and beta diversity were analyzed to understand variations in species composition across regions and time. The study found that land-cover type is a stronger predictor of bird diversity than individual continuous variables such as NDVI or temperature alone. Urbanization, measured by ALAN, exhibits nuanced scale-dependent effects, supporting high abundances of a few generalist species while reducing overall richness. The findings provide actionable insights into the patterns and drivers of avian diversity in Sri Lanka, offering a scalable and reproducible framework for biodiversity research and conservation planning. 

\end{abstract}


\section{Introduction}
Understanding how environmental and anthropogenic factors shape biodiversity is a central challenge in ecology, particularly in regions undergoing rapid land-use change \cite{ref20}. Birds are widely recognized as effective bioindicators due to their sensitivity to habitat structure, climate variability, and human disturbance \cite{ref21}. Sri Lanka combines high ecological heterogeneity within a small geographic extent with increasing pressures from urbanization, agriculture, and climate variability \cite{ref1}. However, large-scale, spatially explicit assessments of bird species richness at the national scale remain limited.

Recent advances in remote sensing and open-access biodiversity data provide new opportunities to address this gap. Satellite-derived indicators such as NDVI, land-cover classifications, and nighttime lights enable consistent characterization of ecological conditions, while citizen-science platforms such as eBird offer extensive species occurrence records \cite{ref5,ref9}. Although these data sources are increasingly available, integrated analyses that jointly evaluate climatic, habitat, and anthropogenic associations with bird diversity remain scarce for tropical island systems.

This study addresses three research questions and tests the following hypotheses:
\begin{itemize}
    \item \textbf{RQ1:} Do categorical habitat variables explain species richness more effectively than continuous environmental variables?
    \item \textbf{H1:} Land-cover type explains cell-level species richness better than continuous environmental variables (NDVI, climate) alone, as measured by higher model rank agreement.
    \item \textbf{RQ2:} Is artificial light at night (ALAN), as a proxy for urbanization, associated with reduced diversity and increased biotic homogenization?
    \item \textbf{H2:} Urbanization (ALAN) is associated with reduced community evenness at the district scale.
    \item \textbf{RQ3:} Are observed richness--environment associations consistent across spatial resolutions?
    \item \textbf{H3:} Species richness summaries and environmental associations remain stable across 2, 5, and 10~km grid resolutions after spatial thinning.
\end{itemize}

This work differs from prior biodiversity studies in three concrete ways: (i)~it provides one of the first nationwide, multi-driver analyses of bird species richness for Sri Lanka; (ii)~it compares environmental, habitat, and urban drivers within the same modelling pipeline rather than in isolation; and (iii)~it evaluates association stability across three spatial scales while combining citizen-science observations with remote sensing, pollution, and climate data.

Citizen-science data introduce known sampling biases. We apply spatial thinning, grid-based aggregation, and effort-corrected temporal metrics so that inferred patterns are interpreted as associations under real-world data constraints rather than as complete censuses. \hf{\dataurl}{Data} and \gh{\codeurl}{code} for this work are publicly available.

\section{Literature Review}
Research on bird diversity has shifted from single-factor explanations toward multi-source, data-driven approaches \cite{ref20,ref21}, but prior work differs in scope and limitations in ways that motivate the present study.

Prior studies have combined satellite imagery with bird records to identify spatial drivers of diversity, emphasizing NDVI and climate productivity \cite{ref1}. That approach demonstrates the value of remote sensing for biodiversity assessment, but it does not systematically compare categorical land-cover structure against continuous greenness when both are available in the same model.
Integrated analyses in Mediterranean rural landscapes have linked remotely sensed indicators with open-access occurrence data, improving spatial coverage for pattern description \cite{ref2}. Such work often prioritizes mapped biodiversity patterns over explicit comparison of habitat versus anthropogenic driver families, and it does not evaluate scale sensitivity.

NDVI is widely used as a proxy for ecological responses to environmental change \cite{ref4}. A key limitation for richness modelling is that NDVI measures photosynthetic greenness rather than structural habitat heterogeneity; identical NDVI values can correspond to forests, croplands, or managed parks with different avian communities.
Agricultural intensification and habitat simplification are associated with farmland bird declines in Europe, highlighting the importance of land-cover context \cite{ref10}. That field-based regional focus differs from nationwide, remotely sensed integration in data-limited tropical islands.

Citizen-science and remote-sensing platforms increasingly support species distribution and forecasting applications \cite{ref5,ref6}. These methods target occurrence prediction accuracy rather than comparing the relative explanatory contribution of habitat structure, climate, pollution, and urbanization within one associational framework.
Mechanistic reviews describe how urbanization and light pollution can alter communities through filtering toward generalists and disrupted phenology \cite{ref12,ref13}. Empirical tests at multiple spatial scales with explicit effort correction remain limited.
Avian diversity changes along urban land-use gradients show that community evenness can decline even when total species counts remain high \cite{ref15}. That pattern motivates district-level dominance analyses, but prior work did not link ALAN, aerosols, and climate in a single harmonized national dataset.
Two gaps persist across this literature: few studies compare habitat structure, environmental conditions, and anthropogenic factors within one pipeline \cite{ref2}, and spatial resolution is rarely treated as an explicit test variable \cite{ref4}. The present study addresses both gaps for Sri Lanka while extending beyond richness to community evenness and compositional turnover \cite{ref14,ref16}.

\section{Methodology}

\subsection{Study Area and Data Sources}

Sri Lanka was selected as the study area due to its exceptional ecological diversity concentrated within a small geographic extent, encompassing tropical lowland rainforests, montane cloud forests, dry-zone scrublands, and coastal wetlands \cite{ref1}. Bird occurrence records were obtained from the Global Biodiversity Information Facility (GBIF) via the eBird citizen-science platform \cite{ref9}, covering the period 2014 - 2024. Environmental datasets include MODIS-derived NDVI and IGBP land-cover classification (17 classes), VIIRS nighttime light radiance as a proxy for urbanization \cite{ref11}, MERRA-2 reanalysis variables for climate and aerosol optical properties, and SRTM elevation data. To construct the final analytical dataset, bird occurrences were merged with environmental and climatic variables by matching geographic coordinates and the month of observation. 
To resolve spatial and temporal mismatches across these multimodal datasets, dynamic environmental features were assigned based on the observation month. Crucially, before modelling, continuous environmental covariates were averaged within spatial grid cells, and categorical variables (like land cover) were assigned their modal class within that cell. This ensured that biological point-occurrences were matched to the broader environmental baseline of their immediate habitat.

To ingest and harmonize the dataset, the eBird observation dataset was established as the primary structural baseline. Relational joins were performed using composite spatio-temporal keys (year, month, longitude, and latitude) for dynamic metrics like NDVI, ALAN, and air quality, while static topographic data such as elevation were integrated using spatial coordinates. 

\subsection{Data Preprocessing and Bias Correction}

Incomplete, duplicate, and unreliable records were removed before the analysis. Continuous variables were assessed for distributional skewness: NDVI and ALAN were normalized using a Yeo-Johnson power transformation \cite{ref22}, while bird abundance was log-transformed for bivariate analyses. Records where the cloud-free coverage band equaled zero were excluded \cite{ref3}. 

To mitigate observer bias inherent in citizen-science data, a grid-based spatial thinning approach was applied: exactly one observation per species, per grid cell, per district. A 5~km grid was chosen as the primary analytical baseline after sensitivity analysis across 2, 5, and 10~km resolutions. The island-wide species inventory (429) remained unchanged across thinning resolutions; this reflects the accumulated species list under thinning rules rather than proof of ecological stability at finer scales. District-level richness gradients were nevertheless similar across 2--10~km grids; 5~km preserved substantially more cells than 10~km while reducing pseudoreplication relative to 2~km.

We did not fit occupancy models with explicit detectability correction, nor did we include eBird checklist duration in the primary richness model. As an effort proxy, we computed per-cell checklist density (thinned record count per grid cell) for sensitivity analyses. Spatial thinning and rarefaction reduce but do not eliminate effort bias.

\subsection{Diversity Metrics and Spatial Aggregation}

Bird diversity was quantified using multiple approaches. Species richness was strictly defined as the discrete count of unique species observed at the cell-level after spatial thinning. This was complemented by presence-based species occupancy (the fraction of thinned grid cells where a species was detected within a district), pairwise Jaccard dissimilarity to measure beta-diversity turnover between districts \cite{ref26}, and relative dominance metrics calculated from district-level summed counts. 

To guarantee that district richness comparisons were fair despite highly uneven sampling effort across regions, an equal-cell rarefaction algorithm was applied \cite{ref17}. A threshold of 300 random subsamples was chosen as the optimal iteration count to stabilize the asymptotic median and 95\% confidence intervals, effectively smoothing the variance caused by differing grid-cell sampling depths.

\subsection{Environmental and Anthropogenic Analysis}

NDVI distributions across land-cover classes were examined using violin plots. Because biological count data violate normality assumptions, Kruskal--Wallis $H$-tests were used to evaluate whether species richness differed across land-cover categories \cite{ref17,ref23}. ALAN served as a proxy for urbanization \cite{ref11}; its associations with MERRA-2 aerosol variables were evaluated using Pearson correlations, while LOWESS regression summarized the non-linear ALAN--NDVI association. False Discovery Rate (FDR) correction controlled false positives in species-level screening \cite{ref18}.

\subsection{Statistical Modelling}

All models were fit on the same district-thinned 5~km cell table ($n = 1{,}736$ cells after land-cover filtering). Driver families were evaluated jointly-habitat (NDVI with modal land cover), climate (weather array), air pollution (aerosol metrics), and urbanization (ALAN with community composition metrics)-before a single integrative model combined representative terms from each family.

We report model skill in three tiers: (i)~\emph{primary}-hold-out Spearman $\rho$; (ii)~\emph{secondary}-hold-out MAE and RMSE; and (iii)~\emph{diagnostic}-McFadden pseudo-$R^2$. The integrative GLM is framed as an exploratory associational model, not a high-accuracy predictive system. We do not use Cox--Snell pseudo-$R^2$.

Modal IGBP class per cell followed the MODIS MCD12Q1 scheme (17 classes globally; 11 classes met the $\geq 25$ cells threshold). Categorical land-cover terms used treatment coding with class 2 (Evergreen Broadleaf Forests) as the reference. Reported contrasts (e.g., class 8 = Woody Savannas, class 12 = Cropland, class 14 = Cropland/Natural Vegetation Mosaics) are interpreted relative to that reference.

Pairwise Spearman correlations among continuous predictors were inspected before modelling. Elevation and temperature were strongly negatively correlated ($\rho \approx -0.87$); precipitation correlated positively with NDVI. The integrative model retained one summary climate term (mean temperature) and one aerosol term. NDVI was always entered with modal land-cover class.

Cell-level species richness is a non-negative integer. A Poisson GLM showed severe overdispersion (Pearson $\chi^2 / \mathrm{df} \approx 38.5$), violating the equidispersion assumption; HC1 standard errors do not correct this distributional mismatch. We therefore used a Negative Binomial GLM as the primary integrative model, which explicitly accommodates extra-Poisson variance. A Poisson specification with HC1 errors was retained only as a sensitivity check; coefficient signs for NDVI, temperature, and land-cover contrasts were qualitatively unchanged. Zero-inflated models were not used because post-thinning cell richness was strictly positive.

ALAN co-varies with observer accessibility; ecological interpretation of urbanization relies on district-level evenness and dominance (H2). A sensitivity model with per-cell checklist density tests whether the ALAN term persists after partial effort adjustment.

Moran's $I$ on Pearson residuals ($k$-nearest-neighbour weights, $k = 8$) quantifies whether GLM independence assumptions are violated.
Model performance was evaluated with a random 80/20 train--test split ($n_{\mathrm{train}} = 1{,}388$, $n_{\mathrm{test}} = 348$, seed 42).

\subsection{Temporal Analysis}

To systematically assess temporal shifts independently of spatial variables and the severe raw-count bias introduced by exponentially increasing observer effort over time, a strictly effort-corrected temporal analysis framework was implemented. Raw observation totals were avoided in favor of three standardized metrics.

\begin{itemize}
    \item Rarefied annual richness, which mathematically forces the identical number of sampled grid cells to be evaluated for each year.
    \item District-year richness normalized per 100 sampled cells.
    \item Temporal species occupancy rates calculated exclusively on a "stable panel" of grid cells defined as cells repeatedly surveyed across a majority of the study years. 
\end{itemize}
Species-specific temporal trends were then estimated using log-linear regression models fitted to these effort-corrected trajectories. Benjamini--Hochberg FDR correction was applied to account for multiple testing across hundreds of species \cite{ref18}. While these methods quantify statistical trends, the unstructured nature of citizen-science data means these slopes represent highly controlled apparent observation trends rather than confirmed absolute population dynamics.

\section{Experiments and Results}

All diversity metrics are computed at the grid-cell level, where species richness represents the number of unique species observed per cell after spatial thinning.

\subsection{Dataset Overview and Spatial Thinning}

After cleaning and integration, the final dataset contained approximately 1.55 million records spanning 25 administrative districts and 429 bird species. Occurrence records were strongly clustered in western and southern coastal areas ($\approx$56\%), with the central ($\approx$26\%), northern ($\approx$16\%), and eastern ($\approx$2\%) regions markedly underrepresented. After thinning at 5~km resolution, the island-wide species inventory remained 429 across 2, 5, and 10~km grids-indicating that the thinned species list is saturated at coarser aggregation, not that cell-level ecology is invariant. Mean district species pool size (243.64 species) was similar across grid scales in the sensitivity analysis.

\subsection{Effects of Habitat and Environmental Variables}
Dual-axis time-series analysis at the national scale indicated that NDVI peaks lag behind seasonal rainfall onset, consistent with vegetation green-up following precipitation pulses.

\begin{figure}[t]
    \centering
    \includegraphics[width=1\linewidth]{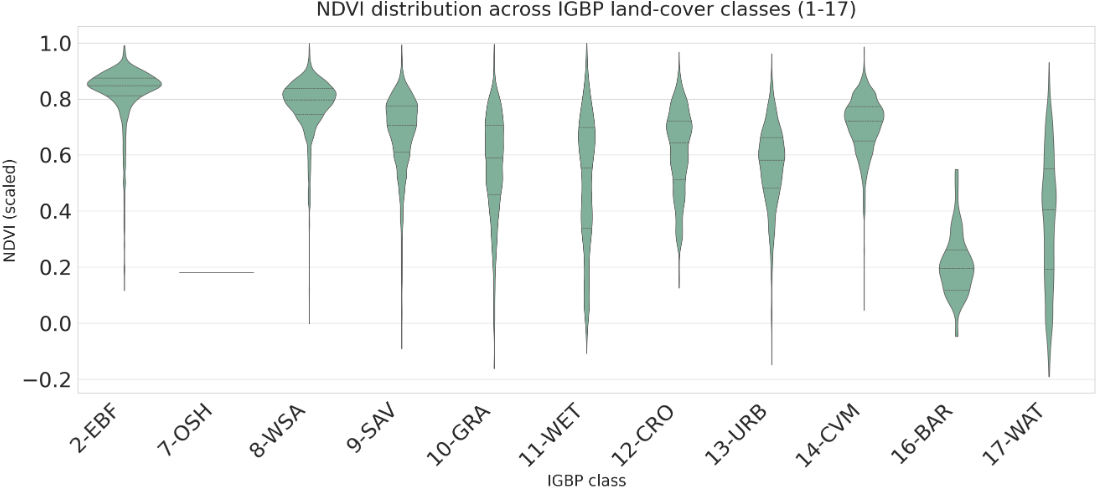}
    \caption{Distribution of NDVI partitioned by IGBP Land Cover classifications, highlighting the stability of evergreen habitats versus the volatility of croplands.}
    \label{fig:NDVI}
\end{figure}

Violin plots of NDVI across IGBP land-cover classes (Fig.~\ref{fig:NDVI}) revealed strong habitat partitioning among the 11 classes with sufficient data. Evergreen forests maintained high and stable NDVI, whereas croplands exhibited pronounced variability. Kruskal--Wallis tests indicated differences in observation counts ($H = 20{,}785$, $p < 0.001$), cell-level species richness ($H = 295$, $p < 0.001$), and Shannon diversity ($H = 333$, $p < 0.001$), with natural habitats supporting higher richness than human-altered landscapes \cite{ref24}. Temperature also varied among districts (Kruskal--Wallis, $p < 0.001$).

Marginal associations with species richness were statistically significant but ecologically small given $n \approx 1.55 \times 10^6$ raw records: NDVI ($\rho = 0.033$, $p < 0.001$), raw counts versus NDVI ($\rho = -0.094$). Land-cover rank association ($\rho = 0.081$) exceeded marginal NDVI associations by a factor of $\approx 2.5$, supporting H1 at the bivariate screening stage. These $\rho$ values denote rank agreement, not variance explained.

\subsection{Effects of Urbanization and Pollution}

\begin{figure}[htbp]
    \centering
    \includegraphics[width=1\linewidth]{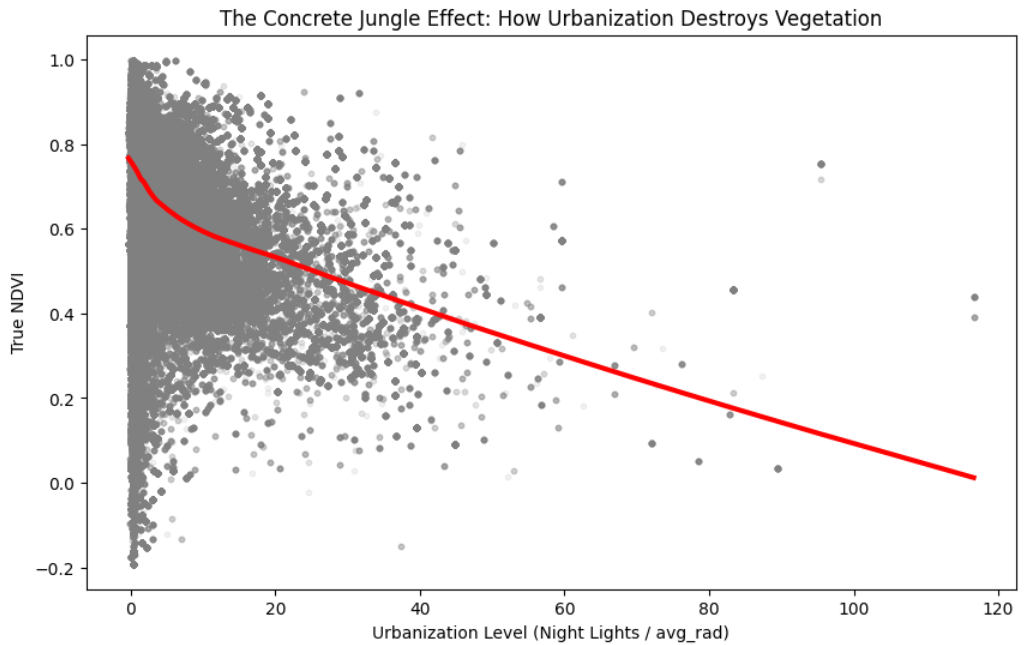}
    \caption{LOWESS regression of NDVI against ALAN, showing lower vegetation greenness in brighter (more urbanized) areas.}
    \label{fig:LOWESS}
\end{figure}

Urbanization, represented by ALAN, showed scale-dependent associations with biodiversity. ALAN correlated positively with aerosol variables, consistent with co-located human activity and pollution. LOWESS regression showed a negative ALAN--NDVI association (Fig.~\ref{fig:LOWESS}), indicating that brighter areas tend to have lower greenness, not necessarily lower richness directly. Overall, ALAN appears associated with community structure at the district scale but not with cell-level richness once sampling effort is considered.

District-level dominance metrics indicated uneven communities in heavily sampled districts, with a few generalist species contributing disproportionately to summed abundance-consistent with H2 regarding reduced evenness and with Berger--Parker and Shannon diversity summaries \cite{ref25,ref24}. Cell-level richness showed a weak positive rank correlation with nighttime radiance (Spearman $\rho \approx 0.11$, $p < 0.001$); this association is reported for completeness but is not interpreted ecologically. Pairwise Jaccard dissimilarity between districts suggested substantial compositional turnover island-wide (Section~\ref{sec:temporal}) rather than district-scale homogenization \cite{ref26}.

LOWESS regression suggested a decline in NDVI with rising ALAN (Fig.~\ref{fig:LOWESS}), consistent with prior evidence of reduced vegetation greenness in brighter areas \cite{ref12}. Pollution showed negligible global richness correlations ($|\rho| \approx 0.01$) but heterogeneous taxon-specific responses (e.g., \textit{Merops persicus}, $|\rho| \approx 0.74$) \cite{ref14}.

\begin{figure}[htbp]
    \centering
    \includegraphics[width=1\linewidth]{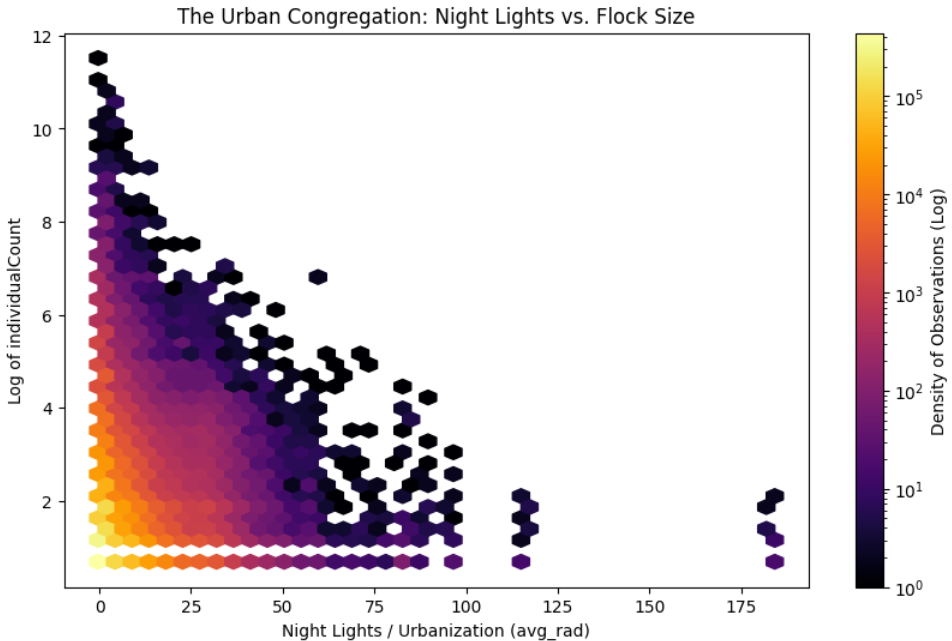}
    \caption{Hexbin density plot of log-transformed per-observation counts against night-light intensity (avg\_rad). Most records occur at low radiance; the largest flock sizes appear at low to moderate light levels, whereas high-radiance areas show fewer observations and generally smaller per-record counts.}
    \label{fig:Hexbin}
\end{figure}

Fig.~\ref{fig:Hexbin} plots log-transformed per-observation counts against night-light intensity. Most records cluster at low radiance; flock sizes are widest there, whereas high-radiance bins are sparse with modest counts-a pattern consistent with uneven observer effort across the urban--rural gradient \cite{ref15}.

\subsection{Multivariate Modelling of Species Richness}

\begin{figure}[htbp]
    \centering
    \includegraphics[width=1\linewidth]{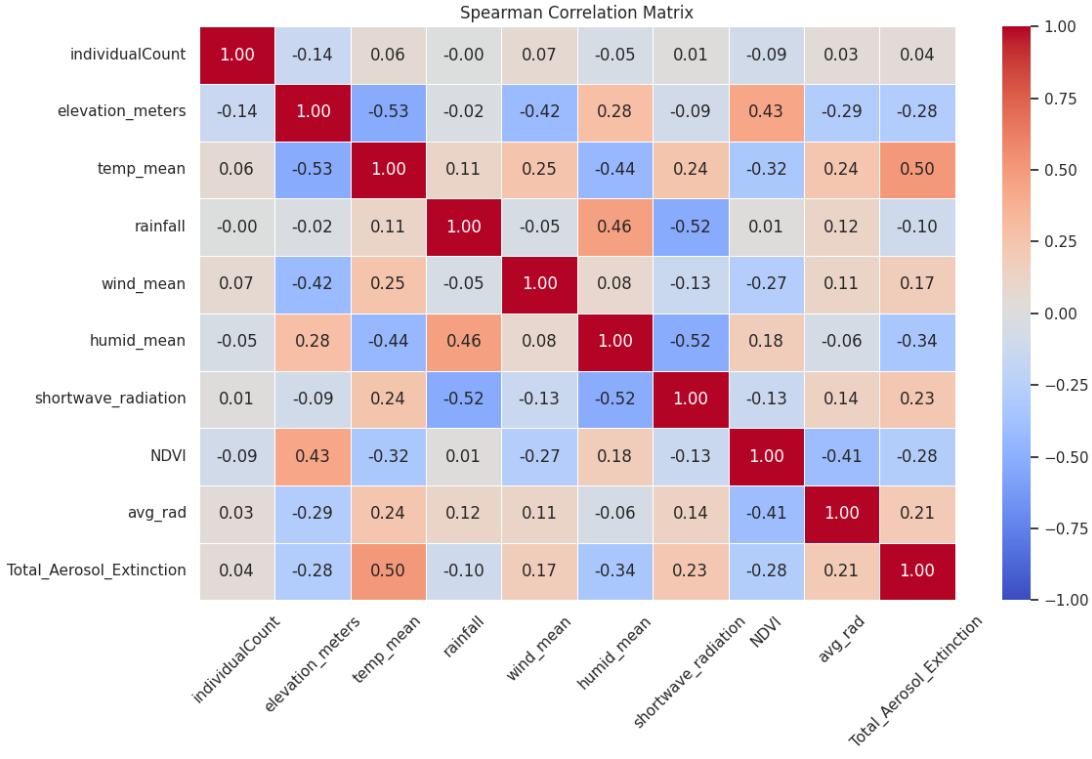}
    \caption{Spearman rank correlation matrix of continuous environmental variables and biological metrics, revealing multicollinearity patterns among predictors.}
    \label{fig:Spearman}
\end{figure}

Fig.~\ref{fig:Spearman} shows multicollinearity among continuous predictors (elevation--temperature $\rho \approx -0.87$; precipitation--NDVI positive). Among driver-family models, climate-only and aerosol-only linear specifications explained little variance ($R^2 = 0.008$ and $0.011$). The joint NDVI + land-cover habitat model achieved the highest rank agreement (Spearman $\rho = 0.150$). The integrative Negative Binomial GLM improved hold-out performance modestly (primary test Spearman $\rho \approx 0.20$; secondary: RMSE $\approx 47$, MAE $\approx 37$ species on $n_{\mathrm{test}} = 348$). McFadden pseudo-$R^2 \approx 0.003$ (diagnostic only) is consistent with an exploratory associational model rather than strong predictive skill.

The integrative Negative Binomial GLM ($n = 1{,}736$ cells, reference land cover = class 2, Evergreen Broadleaf Forests) showed that land-cover contrasts carried most habitat signal relative to NDVI alone. After adjusting for land cover, NDVI had a modest negative partial association ($\beta = -0.390$, SE $= 0.197$, $p = 0.047$, 95\% CI [$-0.776$, $-0.005$]); this sign reversal relative to the marginal $\rho = 0.033$ is consistent with collinearity and model partitioning rather than a global negative NDVI effect. Mean temperature was negatively associated with richness ($\beta = -0.046$, SE $= 0.015$, $p = 0.002$). Elevation, rainfall, and aerosol extinction were not significant ($p > 0.19$). Relative to evergreen forest, woody savannas (class 8, $\beta = -0.187$, $p = 0.005$), cropland (class 12, $\beta = -0.427$, $p = 0.011$), and cropland/natural mosaics (class 14, $\beta = -0.238$, $p = 0.005$) were associated with lower expected richness.

Nighttime radiance entered the integrative model with a positive coefficient ($\beta = 0.066$, $p < 0.001$). When per-cell checklist density was added, the ALAN term fell to $\beta = 0.009$ ($p = 0.60$) while checklist density remained strongly positive ($p < 0.001$), indicating that the raw ALAN term primarily captures sampling intensity rather than ecological richness. Urbanization is therefore discussed using district-level evenness and dominance metrics.

Exploratory Moran's $I$ on GLM Pearson residuals was positive ($I = 0.28$, permutation $p = 0.005$). Spatial autoregressive and occupancy models were not fit in the primary pipeline.

\subsection{Temporal and Regional Diversity Patterns}
\label{sec:temporal}
Monthly species richness and Shannon diversity showed strong seasonality, with higher richness in January--March and November--December and lower Shannon diversity during the late monsoon season \cite{ref24}. Cross-correlation analysis indicated richness aligns more clearly when rainfall leads by several months, consistent with a delayed ecological response through vegetation dynamics \cite{ref4}. Effort-corrected rarefied richness and district-year occupancy rates indicated that temporal trends are not driven purely by sampling variation. 

\begin{figure}[htbp]
    \centering
    \includegraphics[width=1\linewidth]{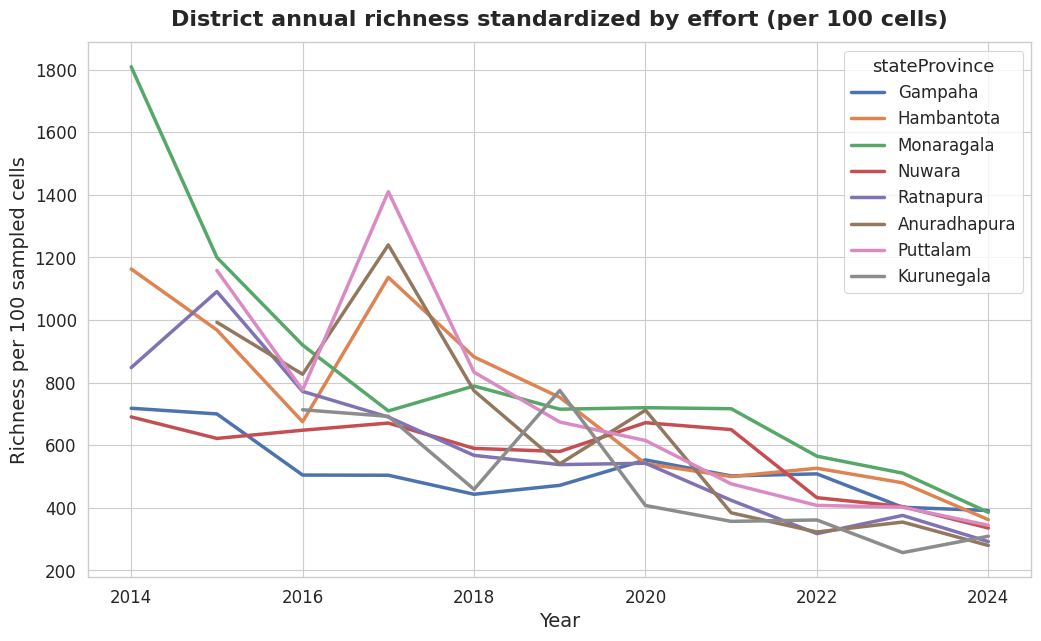}
    \caption{District annual richness standardized by effort (per 100 sampled cells) from 2014 to 2024. After correcting for expanding observer effort, underlying biodiversity trends show a general decline or plateau across the most heavily sampled districts.}
    \label{fig:district_richness}
\end{figure}

As demonstrated in Fig.~\ref{fig:district_richness}, richness standardized per 100 sampled cells shows a plateau or gradual decline in most heavily sampled districts over the past decade. This pattern is consistent with effort-corrected occupancy trends and with declining NDVI in brighter areas, but it cannot be attributed to any single driver without experimental evidence.

Regional comparisons based on pairwise Jaccard dissimilarity suggested substantial differences in community composition across districts. Species-specific log-linear abundance trends indicated non-uniform diversity changes across taxa and regions after Benjamini--Hochberg FDR correction \cite{ref16}.

\section{Discussion}

\subsection{Why land cover outperforms NDVI}
H1 was supported: land-cover type showed stronger associations with species richness than NDVI alone, both marginally ($\rho = 0.081$ versus $0.033$) and in joint models (habitat-module $\rho = 0.150$ versus climate $R^2 = 0.008$). Ecologically, IGBP land-cover classes encode structural habitat differences that NDVI cannot capture. Evergreen forests and woody savannas provide vertical vegetation strata, nesting substrates, and stable food resources year-round in tropical Sri Lanka, whereas croplands and built surfaces may show similar greenness pulses during crop seasons without offering equivalent refugia \cite{ref8,ref10}. NDVI integrates canopy photosynthesis but not canopy height, understory complexity, or land-use history.

The negative partial NDVI coefficient after land-cover adjustment should be read cautiously: it may reflect residual collinearity, seasonal greenness within fixed habitat classes, or model misspecification rather than a biological penalty of greenness. It does not overturn the weak positive marginal association ($\rho = 0.033$) and supports interpreting NDVI only jointly with land cover.

\subsection{Urbanization and community structure}
H2 was partially supported. District-level dominance metrics associated ALAN-rich districts with uneven communities and reduced evenness, consistent with urban filtering described in prior work \cite{ref13,ref15,ref12}. ALAN affects community structure at the district scale but not cell-level richness after effort correction; the cell-level GLM coefficient attenuated when checklist-density proxies were added.

\subsection{Scale stability and model performance}
H3 was partially supported in terms of aggregated richness stability under thinning, though this likely reflects species inventory saturation rather than true ecological invariance across scales. Hold-out Spearman $\rho \approx 0.20$ indicates modest generalization for an exploratory model; McFadden pseudo-$R^2 \approx 0.003$ contrasts with misleadingly high Cox--Snell values for overdispersed count models.

\subsection{Limitations}
Moran's $I = 0.28$ ($p = 0.005$) on residuals indicated positive spatial autocorrelation, so GLM $p$-values may be optimistic \cite{ref19}. Severe Poisson overdispersion ($\chi^2/\mathrm{df} \approx 38.5$) motivated the Negative Binomial primary model; spatial autoregressive or occupancy models with detectability correction remain future priorities. Citizen-science records lack checklist duration and complete effort covariates. Pollution effects on richness were heterogeneous at the taxon level rather than uniform across species \cite{ref14}.

\subsection{Conservation implications}
Land-cover class showed the strongest interpretable associations: woody savannas, cropland, and mixed cropland/natural mosaics were associated with lower richness than evergreen broadleaf forest (reference class). Protecting evergreen forests and woody savannas is plausibly associated with higher species richness at the landscape scale, though causal effectiveness requires intervention studies. Substantial beta diversity across districts suggests region-specific conservation priorities \cite{ref16}.

\section{Conclusion}

We linked eBird records with satellite environmental data to test associations between habitat structure, climate, pollution, urbanization, and bird species richness across Sri Lanka. Land-cover type showed stronger associations than NDVI or climate alone. ALAN was associated with reduced evenness at the district scale but not with cell-level richness after effort correction. The Negative Binomial GLM addresses severe overdispersion and is intended for exploratory inference (hold-out Spearman $\rho \approx 0.20$), not high-accuracy prediction.

\section{Acknowledgments}
The authors acknowledge the Global Biodiversity Information Facility (GBIF) for providing open-access bird occurrence data, and eBird for the citizen-science observation platform. Environmental data were sourced from MODIS, MERRA-2, VIIRS, and SRTM satellite and reanalysis products. The authors also thank the scientific community whose open-source tools and published methodologies informed the analytical framework of this study.

{\footnotesize
\bibliographystyle{IEEEtranN}
\bibliography{references}

@article{ref1,
  author = {Hunt, M. L. and Blackburn, G. A. and Siriwardena, G. M. and Rowland, C. S.},
  title = {Using satellite data to assess spatial drivers of bird diversity},
  journal = {Remote Sens. Ecol. Conserv.},
  volume = {9},
  number = {4},
  pages = {483--500},
  year = {2023},
  doi = {10.1002/rse2.322}
}

@article{ref2,
  author = {Ribeiro, I. and Proenca, V. and Serra, P. and Palma, J. and Domingo-Marimon, C. and Pons, X. and Domingos, T.},
  title = {Remotely sensed indicators and open-access biodiversity data to assess bird diversity patterns in Mediterranean rural landscapes},
  journal = {Sci. Rep.},
  volume = {9},
  pages = {6826},
  year = {2019},
  doi = {10.1038/s41598-019-43330-3}
}

@article{ref3,
  author = {Elvidge, C. D. and Zhizhin, M. and Ghosh, T. and Hsu, F.-C. and Taneja, J.},
  title = {Annual Time Series of Global VIIRS Nighttime Lights Derived from Monthly Averages: 2012 to 2019},
  journal = {Remote Sensing},
  volume = {13},
  number = {5},
  pages = {922},
  year = {2021},
  doi = {10.3390/rs13050922}
}

@article{ref4,
  author = {Pettorelli, N. and Vik, J. O. and Mysterud, A. and Gaillard, J.-M. and Tucker, C. and Stenseth, N. C.},
  title = {Using the satellite-derived NDVI to assess ecological responses to environmental change},
  journal = {Trends Ecol. Evol.},
  volume = {20},
  number = {9},
  pages = {503--510},
  year = {2005},
  doi = {10.1016/j.tree.2005.05.011}
}

@article{ref5,
  author = {Teng, M. and others},
  title = {{SatBird: Bird Species Distribution Modeling with Remote Sensing and Citizen Science Data}},
  journal = {arXiv preprint arXiv:2311.00936},
  year = {2023}
}

@article{ref6,
  author = {Akande, H. A. and Gidado, A. A.},
  title = {{EcoCast: A Spatio-Temporal Model for Continual Biodiversity and Climate Risk Forecasting}},
  journal = {arXiv preprint arXiv:2512.02260},
  year = {2025}
}

@article{ref8,
  author = {Sharma, P. and others},
  title = {Integrating field- and remote sensing data to perceive species heterogeneity across a climate gradient},
  journal = {Sci. Rep.},
  volume = {13},
  pages = {22698},
  year = {2023},
  doi = {10.1038/s41598-023-50812-y}
}

@article{ref9,
  author = {Sullivan, B. L. and others},
  title = {The eBird enterprise: An integrated approach to development and application of citizen science},
  journal = {Biol. Conserv.},
  volume = {169},
  pages = {31--40},
  year = {2014},
  doi = {10.1016/j.biocon.2013.11.003}
}

@article{ref10,
  author = {Donald, P. F. and Green, R. E. and Heath, M. F.},
  title = {Agricultural intensification and the collapse of Europe’s farmland bird populations},
  journal = {Proc. R. Soc. B},
  volume = {268},
  pages = {25--29},
  year = {2001},
  doi = {10.1098/rspb.2000.1325}
}

@article{ref11,
  author = {Elvidge, C. D. and Baugh, K. and Zhizhin, M. and Hsu, F.-C. and Ghosh, T.},
  title = {VIIRS night-time lights},
  journal = {Int. J. Remote Sens.},
  volume = {38},
  number = {21},
  pages = {5860--5879},
  year = {2017},
  doi = {10.1080/01431161.2017.1342050}
}

@article{ref12,
  author = {Gaston, K. J. and Bennie, J. and Davies, T. W. and Hopkins, J.},
  title = {The ecological impacts of nighttime light pollution: A mechanistic appraisal},
  journal = {Biol. Rev.},
  volume = {88},
  number = {4},
  pages = {912--927},
  year = {2013},
  doi = {10.1111/brv.12036}
}

@article{ref13,
  author = {McKinney, M. L.},
  title = {Urbanization, Biodiversity, and Conservation},
  journal = {BioScience},
  volume = {52},
  number = {10},
  pages = {883--890},
  year = {2002},
  doi = {10.1641/0006-3568(2002)052[0883:UBAC]2.0.CO;2}
}

@article{ref14,
  author = {McGill, B. J. and others},
  title = {Species abundance distributions: moving beyond single prediction theories to integration within an ecological framework},
  journal = {Ecol. Lett.},
  volume = {10},
  number = {10},
  pages = {995--1015},
  year = {2007},
  doi = {10.1111/j.1461-0248.2007.01094.x}
}

@article{ref15,
  author = {Blair, R. B.},
  title = {Land Use and Avian Species Diversity Along an Urban Gradient},
  journal = {Ecol. Appl.},
  volume = {6},
  number = {2},
  pages = {506--519},
  year = {1996},
  doi = {10.2307/2269387}
}

@article{ref16,
  author = {McCain, C. M.},
  title = {Global analysis of bird elevation diversity},
  journal = {Global Ecol. Biogeogr.},
  volume = {18},
  number = {3},
  pages = {346--360},
  year = {2009},
  doi = {10.1111/j.1466-8238.2008.00443.x}
}

@article{ref17,
  author = {Gotelli, N. J. and Colwell, R. K.},
  title = {Quantifying biodiversity: procedures and pitfalls in the measurement and comparison of species richness},
  journal = {Ecol. Lett.},
  volume = {4},
  number = {4},
  pages = {379--391},
  year = {2001},
  doi = {10.1046/j.1461-0248.2001.00230.x}
}

@article{ref18,
  author = {Benjamini, Y. and Hochberg, Y.},
  title = {Controlling the false discovery rate: a practical and powerful approach to multiple testing},
  journal = {J. R. Stat. Soc. B},
  volume = {57},
  number = {1},
  pages = {289--300},
  year = {1995}
}

@article{ref19,
  author = {White, H.},
  title = {A heteroskedasticity-consistent covariance matrix estimator and a direct test for heteroskedasticity},
  journal = {Econometrica},
  volume = {48},
  number = {4},
  pages = {817--838},
  year = {1980}
}

@article{ref20,
  title={Increasing awareness of avian ecological function},
  author={Sekercioglu, Cagan H},
  journal={Ecology Letters},
  year={2006}
}

@article{ref21,
  title={Modelling the European Farmland Bird Indicator in response to forecast land-use change in Europe},
  author={Scholefield, Paul and Firbank, Les and Butler, Simon and Norris, Ken and Jones, Laurence M and Petit, Sandrine},
  journal={Ecological Indicators},
  volume={11},
  number={1},
  pages={46--51},
  year={2011},
  doi={10.1016/j.ecolind.2009.09.008}
}

@article{ref22,
  title={A new family of power transformations to improve normality or symmetry},
  author={Yeo, In-Kwon and Johnson, Richard A},
  journal={Biometrika},
  volume={87},
  number={4},
  pages={954--959},
  year={2000},
  doi={10.1093/biomet/87.4.954}
}

@article{ref23,
  title={Use of ranks in one-criterion variance analysis},
  author={Kruskal, William H. and Wallis, W. Allen},
  journal={Journal of the American Statistical Association},
  volume={47},
  number={260},
  pages={583--621},
  year={1952},
  doi={10.1080/01621459.1952.10483441}
}

@article{ref24,
  title={A mathematical theory of communication},
  author={Shannon, Claude E},
  journal={Bell System Technical Journal},
  volume={27},
  pages={379--423, 623--656},
  year={1948}
}

@article{ref25,
  title={Diversity of planktonic foraminifera in deep-sea sediments},
  author={Berger, W H and Parker, F L},
  journal={Science},
  volume={168},
  number={3937},
  pages={1345--1347},
  year={1970},
  doi={10.1126/science.168.3937.1345}
}

@article{ref26,
  title={The distribution of the flora in the alpine zone},
  author={Jaccard, Paul},
  journal={New Phytologist},
  volume={11},
  number={2},
  pages={37--50},
  year={1912}
}
}

\end{document}